\documentclass{kapproc} 

\usepackage{procps} 

\usepackage[dvips]{graphicx}

\kluwerbib 

\begin{document}

\articletitle[Greenberg lecture]
{The cosmic background:\\
Evolution of infrared galaxies and dust properties.\\
A lecture dedicated to the memory of Mayo Greenberg}

\author{Jean-Loup Puget\altaffilmark{1}, Guilaine Lagache\altaffilmark{1} 
and Herve Dole\altaffilmark{1}} 
 
\affil{\altaffilmark{1}Institut d'Astrophysique Spatiale, Universite Paris Sud,F-91405 Orsay, France }

\begin{abstract}
In recent years it became clear that a population of galaxies
radiating most of their power in the far-infrared contribute an
important part of the whole star formation activity in the universe.
These galaxies emit up to 99\% of their energy output in the infrared
by dust. The optical properties of dust in galaxies are thus very
important to understand this population.  Mayo Greenberg who has
brought many of the ideas underlying our understanding of the physics
of interstellar dust has been a pioneer in predicting the important
role of organic solid material formed on dust grains in molecular
clouds and of potential importance of transient heating of very small
particles. It appeared with the {\it ISO} data on galaxies and even more
today with the first {\it Spitzer} data that these mechanisms are important
globally for the observations of infrared galaxies at significant
redshifts. The understanding of their evolution is one of the keys to
the understanding of galaxy built up and evolution.

\end{abstract}

\begin{keywords}
infrared galaxies, cosmic background, galaxy formation
\end{keywords}

\section{Ultra Luminous Infrared Galaxies}

A fraction of the stellar radiation produced in galaxies is absorbed
by dust and re-radiated in the mid and far-infrared. In our Galaxy,
this concerns only about a third of the total luminosity, much less in
elliptical galaxies. The effect of the "dust mask" has been identified
for a long time as an obvious nuisance for optical observations and 
implies a "correction" to account for the total energy
output of galaxies. Now that galaxies have been found for which most of
the radiation is coming out at long wavelengths, the infrared part of
the spectrum cannot be treated as a "correction" to optical
observations.  Although a few very luminous galaxies were observed in
the seventies, it is really with the IRAS survey that a proper census
of the infrared emission of galaxies at low redshift was properly
carried out. The luminosity function at 60 and 100 $\mu m$ is
dominated by $L_\star$ spiral galaxies as could be expected, but a high
luminosity power law tail of luminous galaxies was found. This
population, although not dominant, was carrying a substantial of the
infrared energy production in the local universe which was only one
third of the optical one. It was also found that these luminous
galaxies were often associated with interacting of merging
galaxies. Some were clearly starbursts and others AGNs.

With {\it ISO}, the sensitivity allowed to investigate the more distant
universe up to a redshift of about one and to investigate through
spectroscopic studies whether AGN activity or starburst activity was
powering Ultra Luminous InfraRed Galaxies (ULIRG).

The simplest expectation on the one hand was that heavy elements, and
thus dust, should have decreasing abundances when going to high
redshifts and thus the ratio of infrared to optical should
decrease.The association with mergers, on the other hand, could mean
an increase of the fraction of infrared galaxies in the past due to
the higher number density.

We'll refer to "infrared galaxies" and to "optical galaxies" for short
to mean galaxies in which the infrared emission, respectively
optical, emission dominates. The two populations are rather well
separated (not necessarily meaning that one galaxy does not go from
one class to the other).

\section{Properties of interstellar dust and the spectral energy distribution of infrared galaxies}

For a population where most of the stellar radiation is absorbed and
re-radiated by dust, dust properties and the physics of the absorption
and emission are essential as they determine the Spectral Energy
Distribution (SED) of the galaxy.

Interstellar dust is the topic to which Mayo Greenberg brought a major
contribution. In fact he was really visionary on some topics of the
interstellar dust physics.  Dust grains which contribute most of the
optical extinction must have a size between 0.01 and 0.1 $\mu m$ to
account for the fact that the albedo drops significantly for
wavelengths larger than about 1 $\mu m$. Considering the energy density
of the radiation in a galactic disc like ours, the temperature of a
dust grain is rather low: 15 to 25 K. The emission is peaked at
typically 100 $\mu m$: the wavelength of the emitted radiation is much
larger than the grain size when the absorbed radiation has a
wavelength comparable to the grain size. This has an important
consequence. The emissivity in the far-infrared will decrease roughly
like the square of the wavelength which in turn makes the temperature
dependence on the radiation energy density $u$ very weak ($ T \simeq
u^{1/6}$). For a galaxy like ours the infrared part of the SED peaks at
170 $\mu m$ when for a ULIRG it will peak around 60 $\mu m$: a factor
$3$ in temperature for a factor $10^3$ in energy density or
luminosity. At long wavelengths in the submillimeter and millimeter,
the intensity should decrease like $I_\nu \simeq \lambda^{-4}$. At
wavelengths shorter than this peak the exponential cut off of the
Planck function leaves little power emitted in the mid infrared (5 to
25 $\mu m$), except in regions very close to stars which should carry
only a very small fraction of the integrated luminosity of a galaxy.

The emission spectrum of diffuse cirrus clouds only heated by the
diffuse interstellar radiation field is shown in figure 1
(\cite{boulan1}). The peak emission and the long wavelength behaviour
fit with expectations. On the other hand about one third of the power
is concentrated in the region 5 to 50 $\mu m$ far more than what is
expected from the modified black body at the equilibrium temperature;
this emission contains a set of prominent features between 5 and 18
$\mu m$.  {\it ISO} has shown that many infrared galaxies were showing this
behaviour in their integrated spectra.

Mayo Greenberg (\cite{greenberg1}) proposed a new mode of emission for
interstellar dust grains when they are small enough. The idea of very
small interstellar dust grains has been proposed by Platt (\cite{platt})
earlier but the emission was not discussed at that time. Mayo made
the point that the heat capacity could be small enough and the
probability to absorb a photon small enough that the dust grain could
be transiently heated to high temperature and radiate before absorbing
the next photon.  This mechanism was invoked by K. Sellgren
(\cite{sellgren}) to explain reflection nebulae observations and
\cite{leger} proposed that large Polycyclic Hydrocarbons (PAHs)
molecules could be responsible for the so called "unidentified
infrared bands". \cite{puget1} proposed that this could also explain
the mid infrared emission of the Galactic disc and predicted the
cirrus emission now observed.
 
Mayo Greenberg made another related far reaching prediction: the
grains in dense molecular clouds opaque to optical radiation are very
cold (5 to 10 K). Molecules can condense on these grains in ice
mantles in which chemistry induced by cosmic rays and UV radiation
when the grains reach the surface of the cloud produces larger
molecules and ultimately organic solid material. Experiments in Mayo's
laboratory demonstrated this mechanism and showed the famous "yellow
stuff" residue. We know today that the PAHs are very underabundant in
dense clouds but are overabundant with respect to the diffuse medium
of the surfaces of molecular clouds to disappear again in HII
regions. This is very likely to be the sign of the physics described
by Mayo Greenberg. The organic material seen in absorption at 3.4 $\mu
m$ (aliphatic hydrocarbons, \cite{willner}) is seen in emission at
3.28 $\mu m$ (aromatic hydrocarbons) at the surface of clouds and it
has been shown that aliphatic hydrocarbons transform into aromatic
ones when heated at high enough temperature. We are thus probably
watching directly the mantle evaporation and transformation of the
yellow stuff predicted by Mayo into free large PAH molecules
(\cite{greenberg2}).
\\
The SED of infrared galaxies has thus important properties for the
observations of redshifted ones:
\begin{itemize}
\item
The long wavelength spectrum is very steep ($\simeq \lambda^{-4}$)
\item
A substantial fraction of the energy comes out in a set of features
between 5 and 9~$\mu m$
\item
The SED peaks near 100 $\mu m$ leading to "negative K-corrections" in the
submillimeter
\item
The main PAH feature leads to "negative K-corrections" in the 10 to 30~$\mu m$ range
\end{itemize}
The ``negative K-corrections'' makes redshifted infrared 
galaxies easier to observe. This effect
is particularly strong at long wavelengths: at 1 mm the flux of a
ULIRG radiating $10^{12}$ L$_{\rm \odot}$ is seen as a source with a
constant flux of 1 mJy for $1.5 < z < 10$ !

\section{The Cosmic Infrared Background (CIB)}

The CIB is the energy content of the universe
today produced by galaxies at all redshifts and seen as an isotropic
extragalactic background radiation. It was predicted by Patridge and
Peebles (1967) that observations of such a background gives a powerful
tool to constrain cosmological evolution especially if you observe at
wavelengths larger than a maximum of emission. \cite{partridge} made
that point for the near infrared beyond the peak of the stellar
emission of galaxies ( $\lambda > 1 \mu m$).

The detection of the CIB was the major objective of the DIRBE
experiment aboard COBE. In fact it was by using the FIRAS spectrometer
data that the CIB was first detected at long wavelengths: $\lambda >
200 \mu m$ (\cite{puget2}).  The CIB has also been detected by DIRBE at
2.4, 3.5, 100, 140, 240 $ \mu m$ (see \cite{hauser} for a review). The
Cosmic background is also measured in the optical (\cite{bernstein}).  In
the mid infrared the interplanetary zodiacal dust emission is so
strong that only upper limits were obtained by DIRBE. The combination
of number counts by ISOCAM at 15 $\mu m$ (see \cite{cesar1}) and by
{\it Spitzer} at 24 $\mu m$ (\cite{papo}) with the observations of
TeV gamma ray emission from distant AGNs gives a good measurement of
the background at these wavelengths.  The full cosmic background
spectrum is shown in figure 2 (from \cite{gispert}).

The optical and infrared cosmic backgrounds are well separated and the
first surprising result is that the power in the infrared part is
equal or larger than the power in the optical one although we know
that locally, the infrared output of galaxies is only one third of the
optical. This implies a much stronger evolution of the infrared
luminosity of infrared galaxies than of optical ones.

A second important property to note is that the slope of the long
wavelength part of the CIB: $B_\nu \simeq \lambda^{-2}$ is much less
steep than the long wavelengths spectrum of galaxies. This implies
that the millimeter CIB is not due to the millimeter emission of the
galaxies making the bulk of the emission at the peak of the CIB
($\simeq 170 \mu m$).The millimeter part of the CIB must thus be
dominated by galaxies at rather high redshift for which the SED peak
has been shifted to the sub-millimeter. The millimeter CIB contains
information on the total energy output by these galaxies and about
their spatial distribution which cannot be reached by observation of
individual galaxies before very deep and large surveys are available
at these wavelength which probably requires ALMA.  The implications in
terms of energy output have been drawn by \cite{gispert}. The infrared
production rate per comoving unit volume (1)
 evolves faster between redshift zero and 1 than the optical one
and (2)
 has to stay constant at higher redshifts up to redshift 3 at least.

\section{The luminosity function of infrared galaxies and its evolution}

The first deep ISOCAM surveys in the Hubble Deep Field brought 
spectacular results (see \cite{cesar1} for a review):
\begin{itemize}
\item
 A rather small density of sources detected at 15 $\mu m$ had an
 average brightness which was about 20\% of the optical brightness of
 the galaxies detected by the Hubble Space Telescope
\item
 These infrared galaxies were filling more than half of the CIB at
 this wavelength when compared to the upper limits coming from the
 TeV Gamma rays
\item
Most of these sources were not associated with nearby galaxies
\item
The redshift distribution showed that they are concentrated in the
redshift range 0.5 to 1 with a maximum at 0.8
\item
These sources are high luminosity sources with high infrared to
optical ratio. The average luminosity was $3 \times 10^{11}$L$_{\rm \odot}$.
\end{itemize}
This showed that the infrared output at redshift 0.8 is not dominated
by the infrared part of the SED of the galaxies making the bulk of the
optical output energy but by a small subclass in number of high luminosity
infrared galaxies. These were similar to the local starburst galaxies
detected by IRAS but they dominate the total output. The luminosity
function is dominated by galaxies 30 times more luminous than the
optical $L_\star$ galaxies.  The peak in the redshift distribution
results from the combination of two factors: the strong evolution in
numbers of these galaxies and the negative K-correction due to the
block of PAH features centered around 8 $\mu m$ which peaks at
redshift 0.8 for the 15 $\mu m$ bandpass. The most luminous infrared
galaxies and in some cases the anisotropies of the cosmic background
were also detected by ISO at 170 $\mu m$, SCUBA at 850 $\mu m$ and
IRAM at 1.3 mm. \cite{lagache1} showed that combining
(1) the number counts for the brightest sources and
(2) the CIB spectrum and its anisotropies,
constrains strongly the typical SED of infrared galaxies, the
luminosity function and its evolution with redshift. Figure 3 shows
the typical SED and figure 4 the luminosity function as a function of
redshift. It has a number of implications:
\begin{itemize}
\item
The infrared output energy is dominated by galaxies of luminosity
increasing with redshift from $2 \times 10^{10}$L$_{\rm \odot}$
locally to $3 \times 10^{11}$L$_{\rm \odot}$ at
z=1 and $3 \times 10^{12}$L$_{\rm \odot}$ at redshift larger (2 to 3).
\item
The infrared galaxies are dominated by galaxies with a starburst type
SED.
\end{itemize}
This last conclusion agrees with the spectroscopic studies done with
{\it ISO} by eg. the Garching group who concluded that only 15\% of 
the luminosity is due to AGN activity.

Very recently the first results from the {\it Spitzer} observatory
brought already very interesting new elements on this question
although much more is to be expected in the coming years. One
observational aspect already mentioned above is the negative 
K-correction due to the PAH features. It was predicted by
\cite{lagache1} that, if infrared galaxies keep a similar SED as the
one they have up to redshift 1 up to redshift 2.5 the counts in the 24
$\mu m$ {\it Spitzer}-MIPS band will be affected in a similar way as
the 15 $\mu m$ ISOCAM counts had been. The critical redshift range in
which the main PAH block of features is redshifted in the 24 $\mu m$ 
band is $2 < z < 3$. If the galaxies dominating the luminosity function
are, as predicted, ultraluminous ones with $ L \simeq 3 \times
10^{12}$L$_{\rm \odot}$ this should lead to a maximum in the log N-logS
plot (normalized to the Euclidean one) below 1 mJy. The observations
from Papovich, show clearly a maximum at about 300 $\mu$Jy.  Figure 4
(from \cite{lagache2}) shows how galaxies in different redshift bands
contribute to the counts. The role of the galaxies with redshift
around 2 is clear. This is an indirect indication that the SED has not
changed very much between redshift 1 and 2. This needs to be confirmed
by statistical spectroscopic studies sampling properly the dominant
part of the luminosity function.

\section{Open questions}
If the presence of strong PAH features is confirmed in the typical
spectra of infrared galaxies at redshift beyond 2 which dominate the
energy output at this time, it will shed new light on the chemical
evolution in galaxies and the early production of large amounts of
organic molecules in the universe.

The other open question which is reinforced by the recent {\it Spitzer}
results is related to the luminosity function of infrared
galaxies. When going to larger redshifts the evidence is growing that
the luminosity function is dominated by more luminous galaxies. As
infrared galaxies dominate the energy output at these redshifts, this
shows that the bulk of the star formation takes place in high
luminosity objects. It is unlikely that such large luminosities could
be associated with low mass galaxies. In fact the ISOCAM galaxies had
been showed to be massive ones (\cite{cesar1}).

In the standard model of structure and galaxy formation, dwarf
galaxies are the first ones to collapse at high z and the mass
function of collapsed objects is dominated by low mass galaxies with
an evolution where the mass function is dominated more and more by
massive galaxies as the universe expands. This opposite evolution of
the mass function and of the luminosity function and of the existence
of a class of galaxies with rather small number density contributing
the largest fraction of the energy output has strong implications.\\

The star formation history in galaxies has probably three parts: 

\begin{itemize}
\item
Typically 40\% of the radiation comes from galaxies following a smooth
evolution with the energy output mainly in the optical and near IR,
\item
The second one is the starbust component concentrated in a small
fraction of the lifetime during which the luminosity is 10 to 100
times larger and the output is mostly in the infrared.
\item
The third component is the one associated with the AGN activity which
is probably also dominant only during limited periods but carries only
10 to 20\% of the total radiated energy.
\end{itemize}

This picture is mostly constrained by what we now know on the cosmic
background from galaxies and the identification of its dominant
components.

Infrared galaxies are absent from the largest collapsed structures at
present (galaxy clusters) thus are anti-biased with respect to the
dark matter. They are often associated with mergers and if merging is
a dominant mechanism for galaxy construction, the evolution with time
of infrared galaxies is probably the best tracer we have of this
activity. It is likely that at redshifts larger than 1, infrared
galaxies show a strongly biased distribution with respect to the dark
matter when structures and massive galaxies are just forming as merger
rate should go as the square of the local density.

All of this needs to be investigated by studying the evolution of the
luminosity function and the statistical properties of the spatial
distribution of infrared galaxies. {\it Spitzer} will be great for such
studies but will be strongly limited by confusion for redshift beyond
2-2.5. The study of the structure of the CIB is the tool to study the
spatial distribution of infrared galaxies at larger redshifts with
Planck and Herschel. The ALMA surveys will be needed to get the final
answer to these questions.

\begin{acknowledgments}
It is a pleasure to thanks the sponsors of this
Conference, the Anglo American Chairmans \' Fund, for their financial support of this J.
Mayo Greenberg Lecture
\end{acknowledgments}

\begin{figure}[cirrus]
\vskip.2in
\centerline{\includegraphics[width=5in]{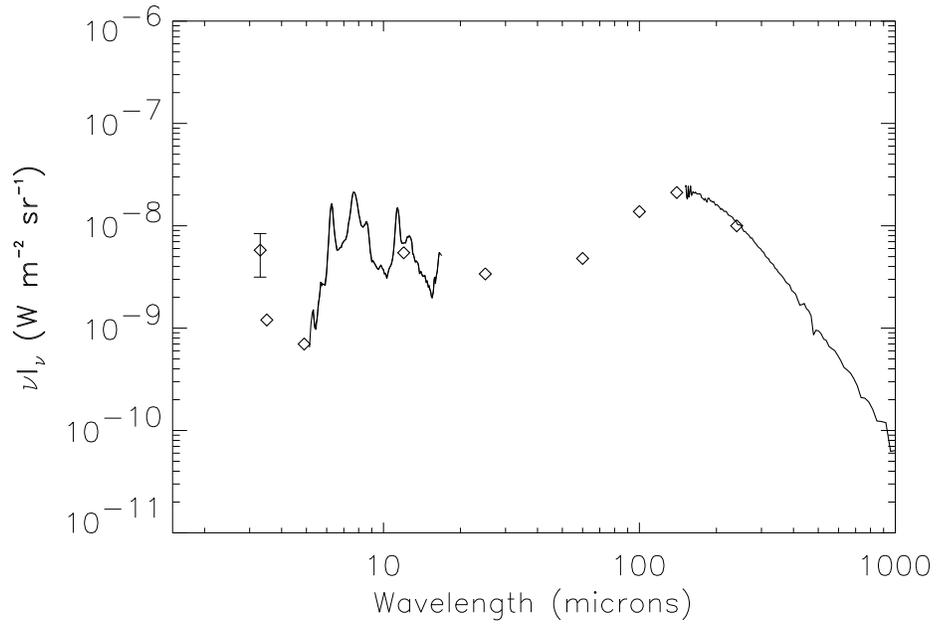}}
\caption{Diffuse interstellar cirrus emission for a column density N(HI)=10$^{20}$ cm$^{-2}$ 
(from \cite{boulan1})}
\end{figure}

\begin{figure}[cib]
\vskip.2in
\centerline{\includegraphics[width=5in]{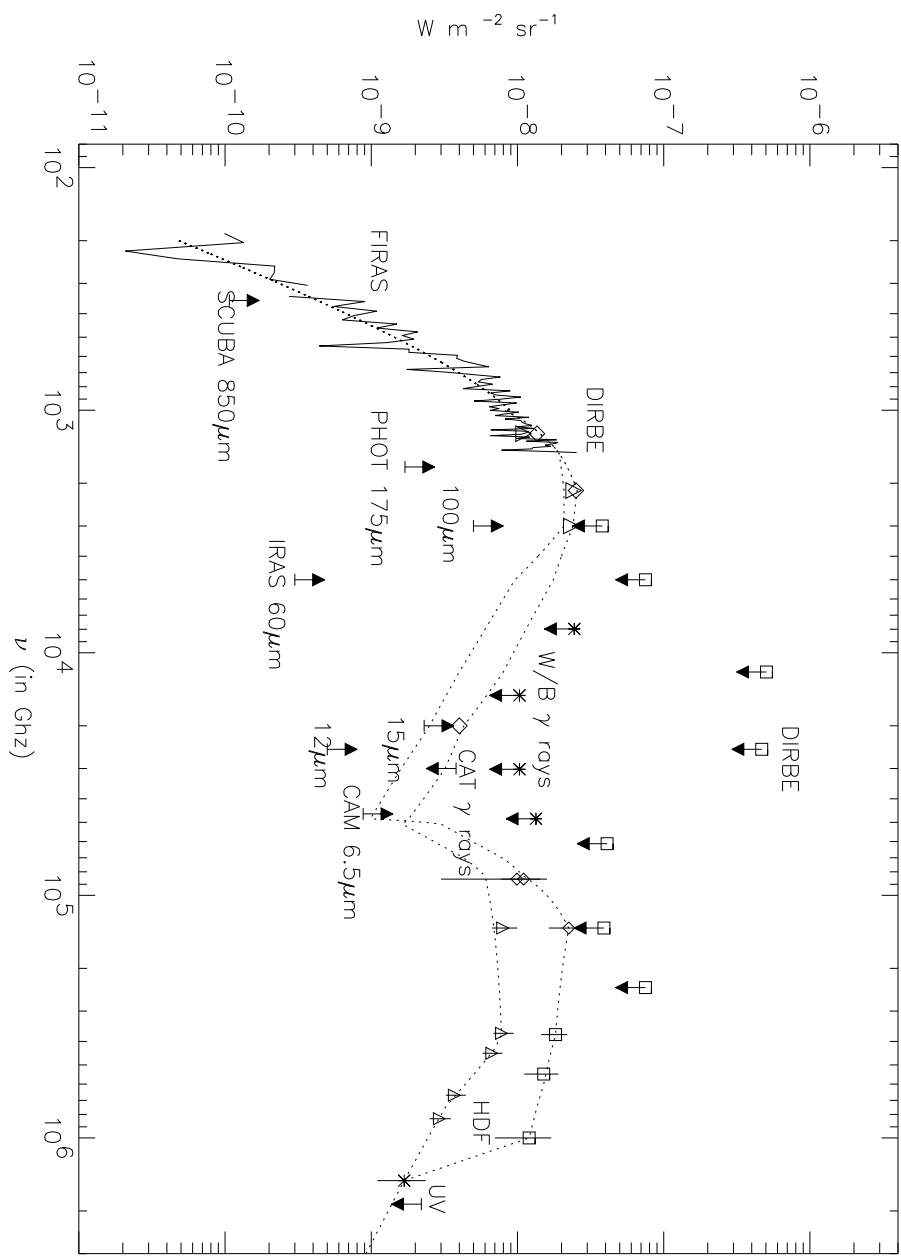}}
\caption{Cosmic Background from the UV to the
millimeter wavelength (from \cite{gispert})}
\end{figure}

\begin{figure}[sed]
\vskip.2in
\centerline{\includegraphics[width=5in]{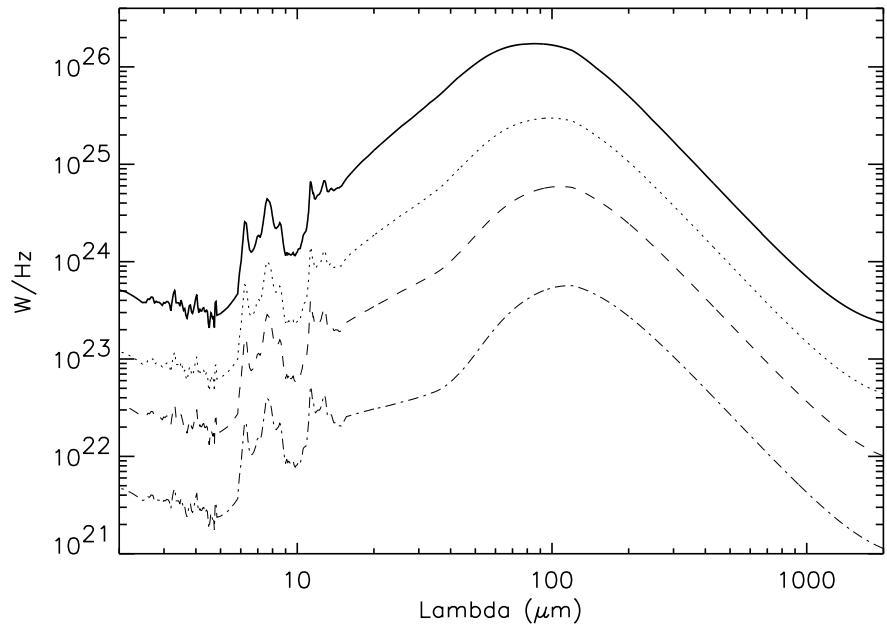}}
\caption{Spectral energy distribution of infrared galaxies.
Starburst model spectra for different luminosities: L=3. 10$^{12}$ L$_{\odot}$
(continuous line), L=5 10$^{11}$ L$_{\odot}$ (dotted line),  L=10$^{11}$ L$_{\odot}$
(dashed line) and  L=10$^{10}$ L$_{\odot}$ (dotted-dashed line). 
From Lagache et al. 2003. }
\end{figure}

\begin{figure}[lfunc]
\vskip.2in
\centerline{\includegraphics[width=6in]{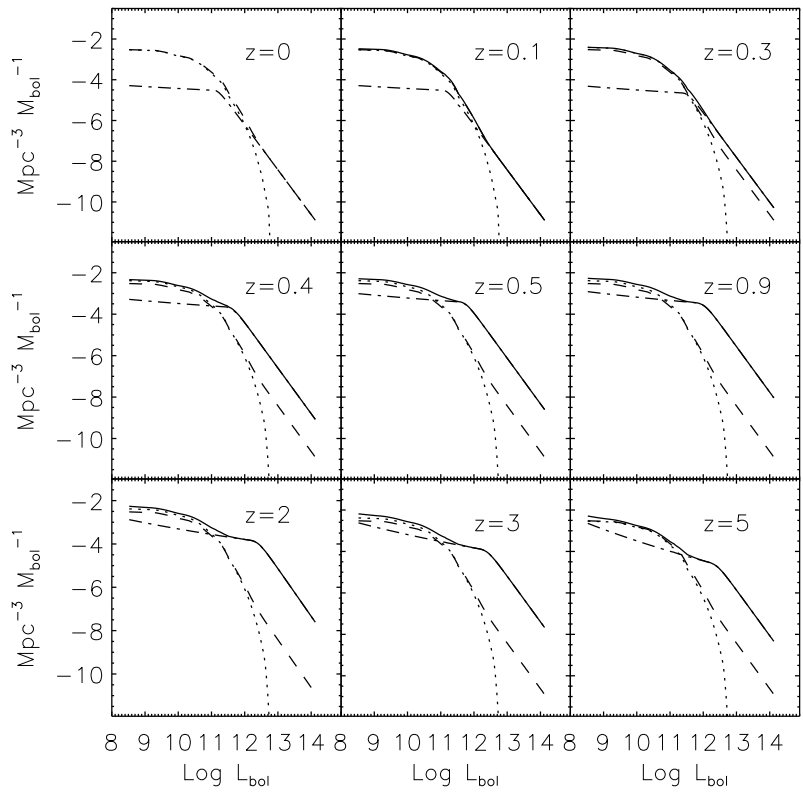}}
\caption{Co-moving evolution of the luminosity function. The dotted line is for the normal
galaxies and the dotted-dashed line, for the starburst galaxies. The continuous
lines corresponds to both starburst and normal galaxies and the dashed line is the LF
at z=0 for comparison (from Lagache et al. 2003).}
\end{figure}

\begin{figure}[24counts]
\vskip.2in
\centerline{\includegraphics[width=4in]{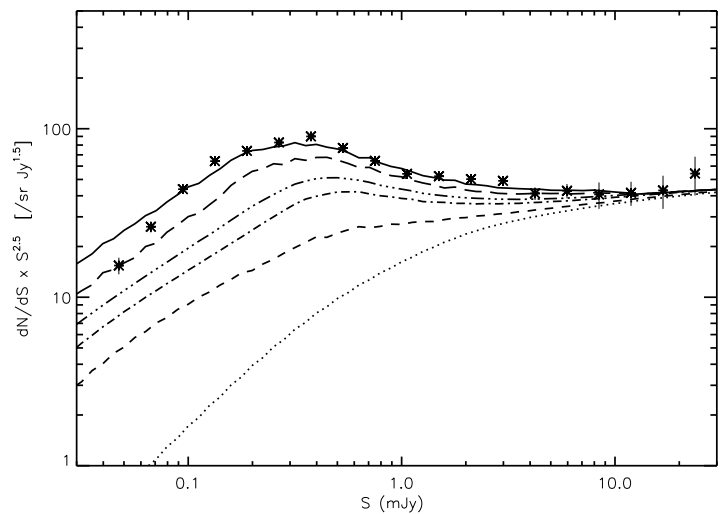}}
\caption{Redshift contribution to the number counts at
24~$\rm \mu$m. The dot, dash, dash-dot, dash-3 dot, long-dash
correspond to the number counts up to redshifts 0.3, 0.8,  1, 1.3 and
2 respectively. The continuous line is the whole redshift contribution.
The model is from Lagache et al. (2004) and
the data are from Papovich et al. (2004).}
\end{figure}

\end{document}